\documentclass[letter,onecolumn]{jpsj3}
\usepackage{txfonts}
\usepackage{color}

\title{Quantum Criticality of Valence Transition for the Unique Electronic State of Antiferromagnetic Compound EuCu$_2$Ge$_2$}

\author{Jun Gouchi$^1$\thanks{gouchi@issp.u-tokyo.ac.jp}, Kazumasa Miyake$^2$, Wataru Iha$^3$, Masato Hedo$^4$, Takao Nakama$^4$, Yoshichika \={O}nuki$^4$, and Yoshiya Uwatoko$^1$}
\inst{$^1$Instittute for Solid State Physics, University of Tokyo, Kashiwa, Chiba 277-8581, Japan \\
$^2$Center for Advanced High Magnetic Field Science, Graduate School of Science, Osaka University, Toyonaka, Osaka 560-0043, Japan\\
$^3$Graduate School of Engineering and Science, University of the Ryukyus, Nishihara, Okinawa 903-0213, Japan \\
$^4$Faculty of Science, University of the Ryukyus, Nishihara, Okinawa 903-0213, Japan} 

\abst{The effect of pressure on the unique electronic state of  the antiferromagnetic (AF) compound EuCu$_2$Ge$_2$ has been measured in a wide temperature range from 10 mK to 300 K by electrical resistivity measurements up to 10 GPa. The N$\mathrm{\acute{e}}$el temperature of $T$$_\mathrm {N}$ = 15 K at ambient pressure increases monotonically with increasing pressure and becomes a maximum of $T$$_\mathrm {N}$ = 27 K at 6.2 GPa but suddenly drops to zero at $P$$_\mathrm {c}$ $\simeq$ 6.5 GPa, suggesting the quantum critical point (QCP) of the valence transition of Eu from a nearly divalent state to that with trivalent weight. The $\rho_\mathrm{mag0}$ and $A$ values obtained from the low-temperature electrical resistivity based on the Fermi liquid relation of $\rho_\mathrm{mag} = \rho_\mathrm{mag0} + AT^{2}$ exhibit huge and sharp peaks around $P$$_\mathrm {c}$. The exponent $n$ obtained from the power law dependence $\rho_\mathrm{mag} = \rho_\mathrm{mag0} + BT^{n}$ is clearly less than 1.5 at $P = P _\mathrm{c}$ $\simeq$ 6. 5 GPa, which is expected at the AF-QCP. These results indicate that $P$$_\mathrm {c}$ coincides with $P$$_\mathrm {v}$, corresponding to the quantum criticality of the valence transition pressure $P$$_\mathrm {v}$. The electronic specific heat coefficient $\gamma$, estimated from the generalized Kadowaki-Woods relation, is about 510 mJ/(mol$\cdot$K$^{2})$ around $P$$_\mathrm {c}$, suggesting the formation of a heavy-fermion state.}

\begin{document}
\maketitle

Most Eu compounds including Eu$_2$Ni$_3$Ge$_5$\cite{1} and EuRhSi$_3$\cite{1} order magnetically because of the divalent electronic state of Eu$^{2+}$. Valence instability often occurs in Eu compounds, from the divalent electronic state of Eu$^{2+}$(4$f^{7}$: $S = $7/2, $L =$ 0, and $J = $7/2) at high temperatures to the nearly nonmagnetic electronic state (4$f^{6}$ in Eu$^{3+}$: $S = L =$ 3, and $J = $0) at low temperatures, upon changing the magnetic field and pressure. Here, $S$, $L$, and $J$ denote the spin, orbital, and total angular momenta, respectively. These compounds can be used to construct a conventional pressure phase diagram\cite{1}. That is very similar to the Doniach phase diagram for the Ce compounds\cite{2, 3, 4}. According to the result of the electrical resistivity measurement of Eu$_2$Ni$_3$Ge$_5$, which is an antiferromagnet with $T_\mathrm {N}$ = 19 K and exhibits a successive change of the antiferromagnetic (AF) structure at $T_\mathrm {N}'$ = 17 K, the Eu valence transition is not realized but the Eu valence changes continuously as a function of pressure\cite{1}. The N$\mathrm{\acute{e}}$el temperature $T_\mathrm {N}$ increases with applying pressure up to 6 GPa, starts to decrease with further increasing pressure, and most likely becomes zero at around $P_\mathrm{c} =$ 12-13 GPa. A similar trend is also realized in EuRhSi$_3$\cite{1}.

In the case of EuRh$_2$Si$_2$, on the other hand, the Eu valence is nearly divalent and EuRh$_2$Si$_2$ orders antiferromagnetically in the low-temperature region. The N$\mathrm{\acute{e}}$el temperature of $T_\mathrm {N}$ = 23.8 K increases as a function of pressure. The change in valence, which is realized in the pressure region of 0.96 GPa $< P <$ 2 GPa, is surprisingly very sharp, indicating the first-order phase transition from nearly divalent to nearly trivalent\cite{5, 6}. With further increasing pressure, its electronic state approaches the trivalent state. EuCo$_2$Ge$_2$ exhibits a similar trend\cite{7}.

Recently, chemical- and high-pressure measurements have been carried out on single crystals of EuCu$_2$(Si$_x$Ge$_{1-x}$)$_2$ grown by the Bridgman method\cite{12}. EuCu$_2$Ge$_2$ with the ThCr$_2$Si$_2$-type tetragonal structure undergoes two successive transitions at $T$$_\mathrm {N}$ $\sim$ 15 K and $T$$_\mathrm {N}' \sim$  9 K, while EuCu$_2$Si$_2$ shows no magnetic orderings at ambient pressure. $T_\mathrm {N}$ slightly increases with increasing $x$ and starts decreasing at $x$ = 0.5, and $T_\mathrm {N}$ disappears around $x$ $\simeq$ 0.7, revealing a mixed-valence or valence-fluctuating state. Similar behavior is observed in Ce-based heavy-fermion compounds\cite{2, 3, 4}. In the case of EuCu$_2$Si$_2$ ($x$ = 1), on the other hand, an intermediate valence state was indicated by resistivity, specific heat, and magnetic susceptibility measurements. EuCu$_2$Si$_2$ is not in the Eu-divalent state but is close to the Eu-trivalent state. These results are consistent with those of previous studies on a series of EuCu$_2$(Si$_x$Ge$_{1-x}$)$_2$\cite{13}. Replacing Ge by Si leads to a chemical compression and indicates the transition from the nearly trivalent state ($x <$ 0.65) to a heavy-fermion state (0.65$< x <$ 0.85) and then to a nonmagnetic state.

In the case of EuCu$_2$Ge$_2$, the effective magnetic moment $\mu_\mathrm {eff}$, obtained from the inverse magnetic susceptibility\cite{12} and  M$\ddot{\mathrm {o}}$ssbauer spectroscopy\cite{14}, is close to the free-ion value $\mu_\mathrm {eff} = $ 7.94$\mu_\mathrm {B}$/Eu of Eu$^{2+}$. $T_\mathrm {N}$ = 15 K at ambient pressure increases as a function of pressure, becomes a maximum of $T_\mathrm {N}$ = 27 K at 6 GPa, and decreases steeply toward zero at a pressure between $P =$ 6 and 7 GPa. Both the $A$ and $\rho$$_0$ values, obtained by assuming the low-temperature resistivity $\rho = \rho_{0} + AT^{2}$ based on the Fermi liquid relation, exhibit pronounced peaks at around $P = $ 7 GPa. The determined value of $A$ is approximately 0.23 $\mu$$\Omega$ cm K$^{-2}$ at 7 GPa. It corresponds to 370 mJ / (mol$\cdot$K$^2$) if we assume the relation of the generalized Kadowaki-Woods plot ($N$ = 4), namely, $A/\gamma^{2} = 2 \times 10^{-5}/N(N-1) \simeq 0.17 \times 10^{-5}$ $\mu$$\Omega$ cm (K mol/mJ)$^{2}$, where $N$ is the degeneracy of $f$-levels\cite{15}. These results suggest that the effective mass of EuCu$_2$Ge$_2$ is highly enhanced at around 7 GPa. From X-ray absorption experiments\cite{16}, the valence continuously changes from 1 to 6 GPa. In the higher-pressure region, the strong magnetic interaction is replaced by the pure valence fluctuations and the valence of Eu increases up rapidly up to 20 GPa. The advantage of a pressure experiment is that it maintains the sample as clean as possible, in contrast to alloying as EuCu$_2$(Si$_x$Ge$_{1-x}$)$_2$. The abrupt drop of $T_\mathrm {N}$ in a very narrow pressure region is highly different from the two cases mentioned above, which are described by the Doniach picture as in Eu$_2$Ni$_3$Ge$_5$\cite{1} and the valence transition as in EuRh$_2$Si$_2$\cite{5, 6}. This suggests that there is a new physical mechanism that cannot be understood within the conventional point of view of the phase change under pressure. However, the pressure ranges in the measurement described in Ref. 8 are not narrow enough considering the sharp disappearance of $T_\mathrm {N}$, indicating the need to perform  measurements with finer intervals of pressure up to 10 GPa and down to much lower temperatures 10 mK. In this paper, we report the results of such resistivity measurements that reveal a new aspect of the transition at $P = P_\mathrm {c}$, where the AF order disappears discontinuously.

Single crystals of EuCu$_2$Ge$_2$ were grown by the Bridgman method using a Mo crucible, as described detail in a previous paper\cite{12}. The electrical resistivity was measured by a conventional ac four-probe method using a Linear Research LR-700 ac resistance bridge. The current flow $J$ was chosen to be parallel to the tetragonal [100] direction. High pressure experiments were performed by using a palm cubic anvil cell, which is known to generate hydrostatic pressure owing to the multiple-anvil geometry\cite{17}. A mixture of Flourinert 70 and 77 were used as a pressure medium. The  pressure was determined using a laboratory-built calibration table, which was calibrated by detecting the characteristic phase transitions of Bi, Sn, and Pb\cite{17}. The pressure cell was set to a dilution refrigerator using BlueFors LD-400 and cooled to 10 mK.

\begin{figure}
\begin{center}
\includegraphics*[width=0.8\linewidth]{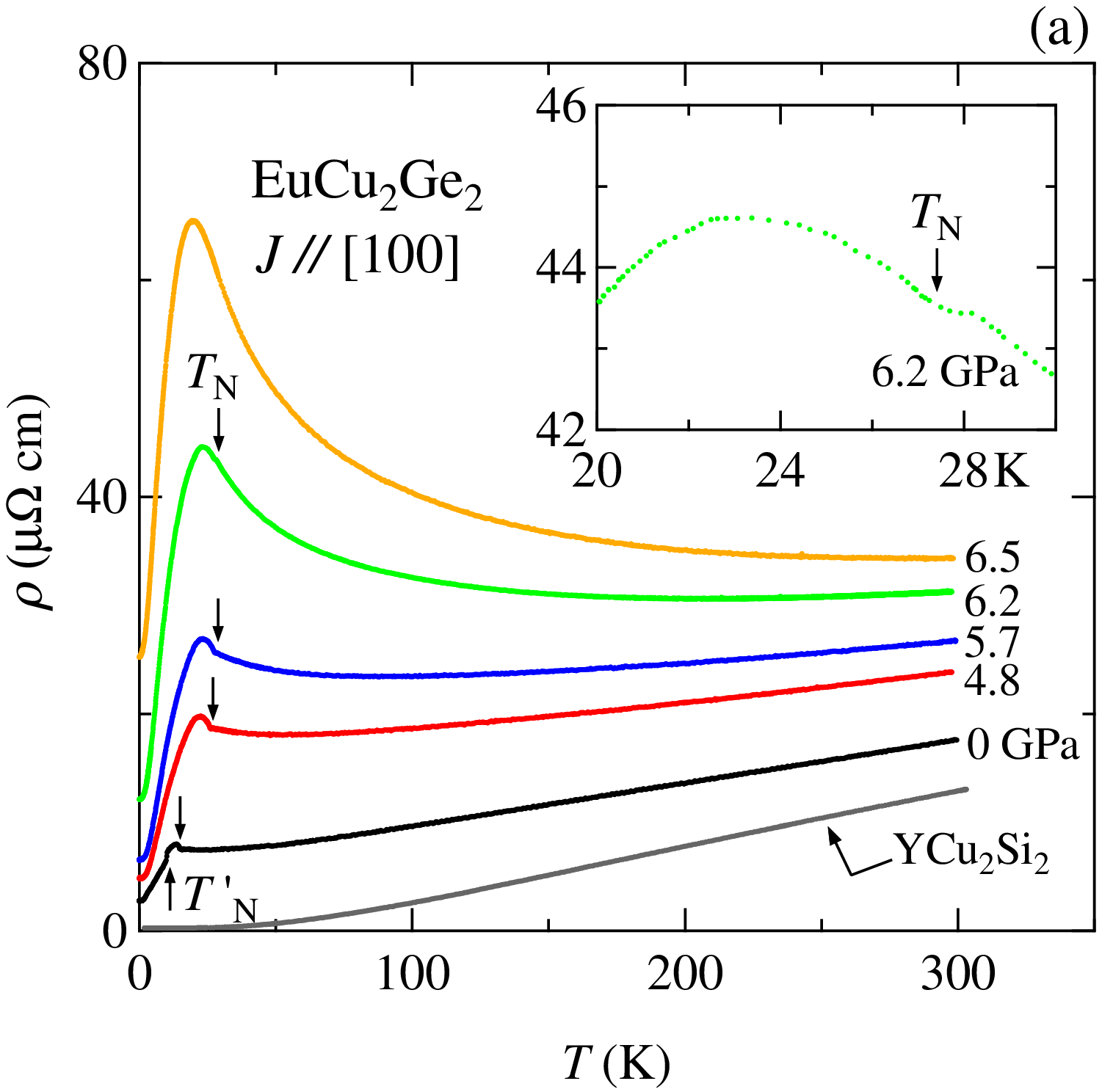}
\includegraphics*[width=0.8\linewidth]{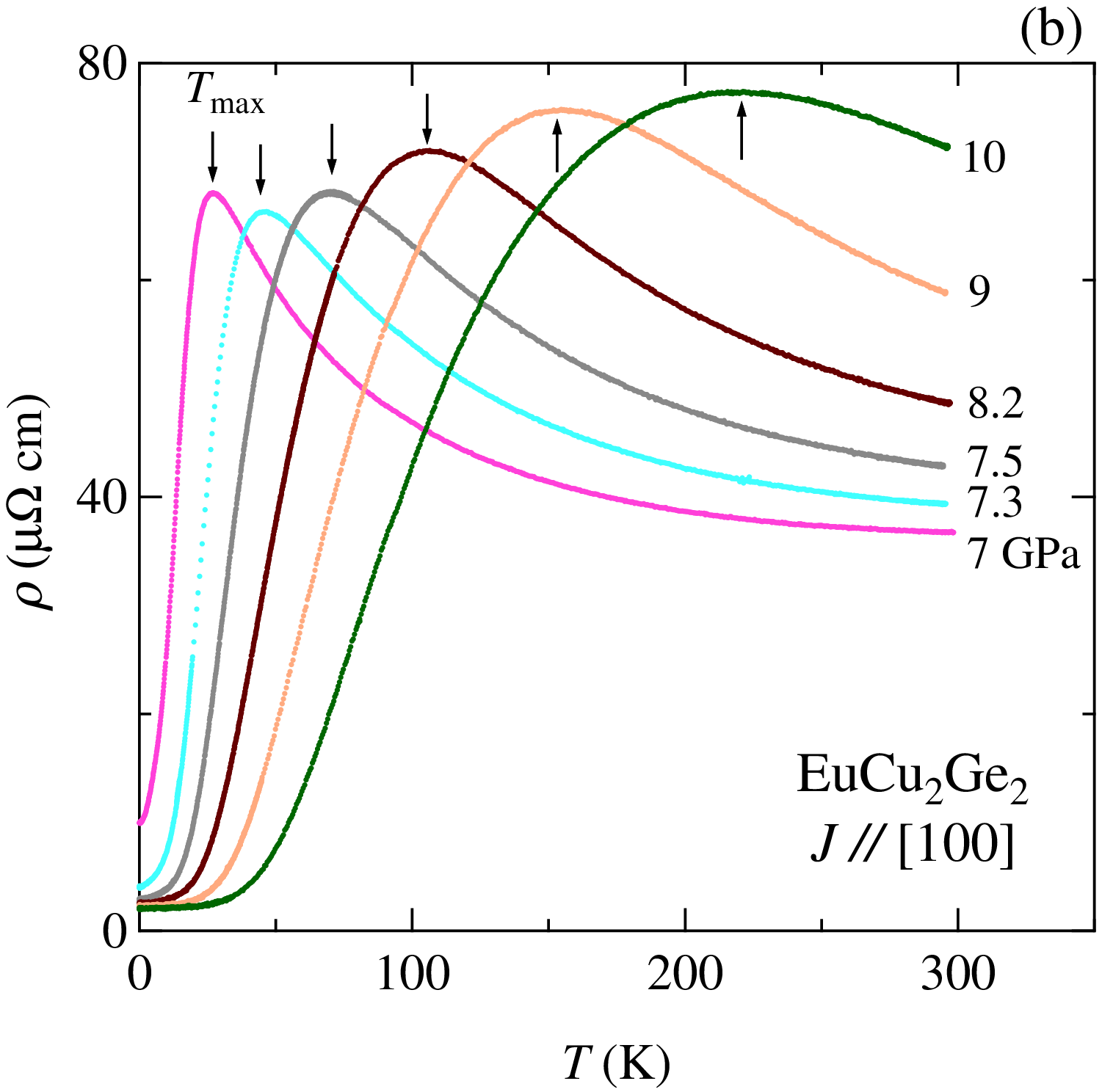}
\caption{\label{fig1} (color online) Temperature dependences of the electrical resistivity in  EuCu$_2$Ge$_2$ (a) up to 6.5 GPa and (b) up to 10 GPa. The arrows indicate (a) the AF transitions and (b) $T$$_\mathrm{max}$. The inset of (a) shows a magnified view of the electrical resistivity at 6.2 GPa.}
\end{center}
\end{figure}
Figures 1(a) and 1(b) show the temperature dependences of the electrical resistivity $\rho$ under several pressures, together with the data for YCu$_2$Si$_2$\cite{18} as a reference of the phonon parts of EuCu$_2$Ge$_2$. At ambient pressure, the electrical resistivity decreases almost linearly with decreasing temperature and becomes constant below 30 K. The step like increase and decrease in resistivity correspond to the AF transition at $T$$_\mathrm {N}$ = 15 K and $T$$_\mathrm {N}'$ = 8.9 K, respectively. With increasing pressure, the resistivity at room temperature increases because the $c$-$f$ hybridization between conduction electrons and 4$f$ electrons is enhanced. $T$$_\mathrm {N}$ substantially increases from 15 K and becomes maximum at about 27 K at 6.2 GPa. Although  $T$$_\mathrm {N}$ at 6.2 GPa is unclear in a wide temperature region, a slight change can be found as shown in the inset of Fig. 1(a). The temperature dependences of $\rho$ in EuCu$_{2}$Ge$_{2}$ are similar to the  resistivity results for Eu$_2$Ni$_3$Ge$_5$\cite{1} and EuCu$_2$(Si$_x$Ge$_{1-x}$)$_2$\cite{12}. They suggest that the valence of Eu increases continuously from nearly divalent to trivalent with increasing pressure. There is no anomaly reflecting the valence transition $T$$_\mathrm {v}$ at low temperatures. These data are similar to the recently reported ones and the temperature dependences of $\rho$ under high pressures are reminiscent of the Kondo effect in Ce-based heavy-fermion compounds\cite{2, 3, 4}. The resistivity at 6.5 GPa indicates no clear magnetic ordering and is typical for a moderate heavy-fermion state. Note that $T$$_\mathrm{N}'$ disappears rapidly under pressure, where it was not observed at 2 GPa in a previous experiment\cite{12}. The temperature dependence of the electrical resistivity at 6.5 GPa resembles that of EuCu$_2$(Si$_x$Ge$_{1-x}$)$_2$ with $x$ = 0.6, where the N$\mathrm{\acute{e}}$el temperature decreases rapidly with increasing concentration $x$. The electrical resistivity possesses a peak around $T_\mathrm{max} =$ 20 K under pressure and it is characteristic of the Kondo behavior, where $T$$_\mathrm{max}$ is defined as the Kondo temperature. The resistivity shows a -log$T$ dependence. $T$$_\mathrm{max}$ remains up to 6.5 GPa and shifts to higher temperatures with further increasing pressure. The qualitative behavior of the electrical resistivity of EuCu$_2$Ge$_2$ under higher pressures will gradually approach that of EuIr$_2$Si$_2$\cite{19}, which is a nonmagnetic compound and exhibits a broad maximum around 200 K in $\rho$($T$).

\begin{figure}
\begin{center}
\includegraphics*[width=0.9\linewidth]{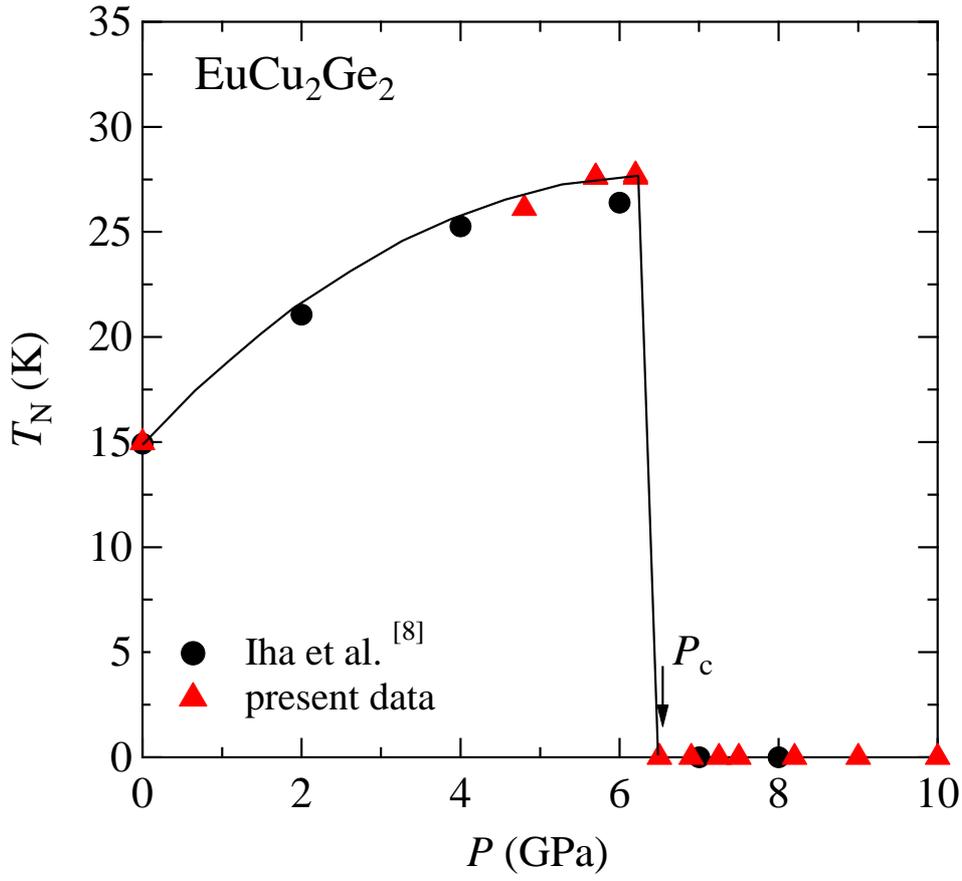}
\caption{\label{fig2} (color online) Pressure-N$\mathrm{\acute{e}}$el temperature phase diagram in EuCu$_2$Ge$_2$.}
\end{center}
\end{figure}
\begin{figure}
\begin{center}
\includegraphics*[width=0.95\linewidth]{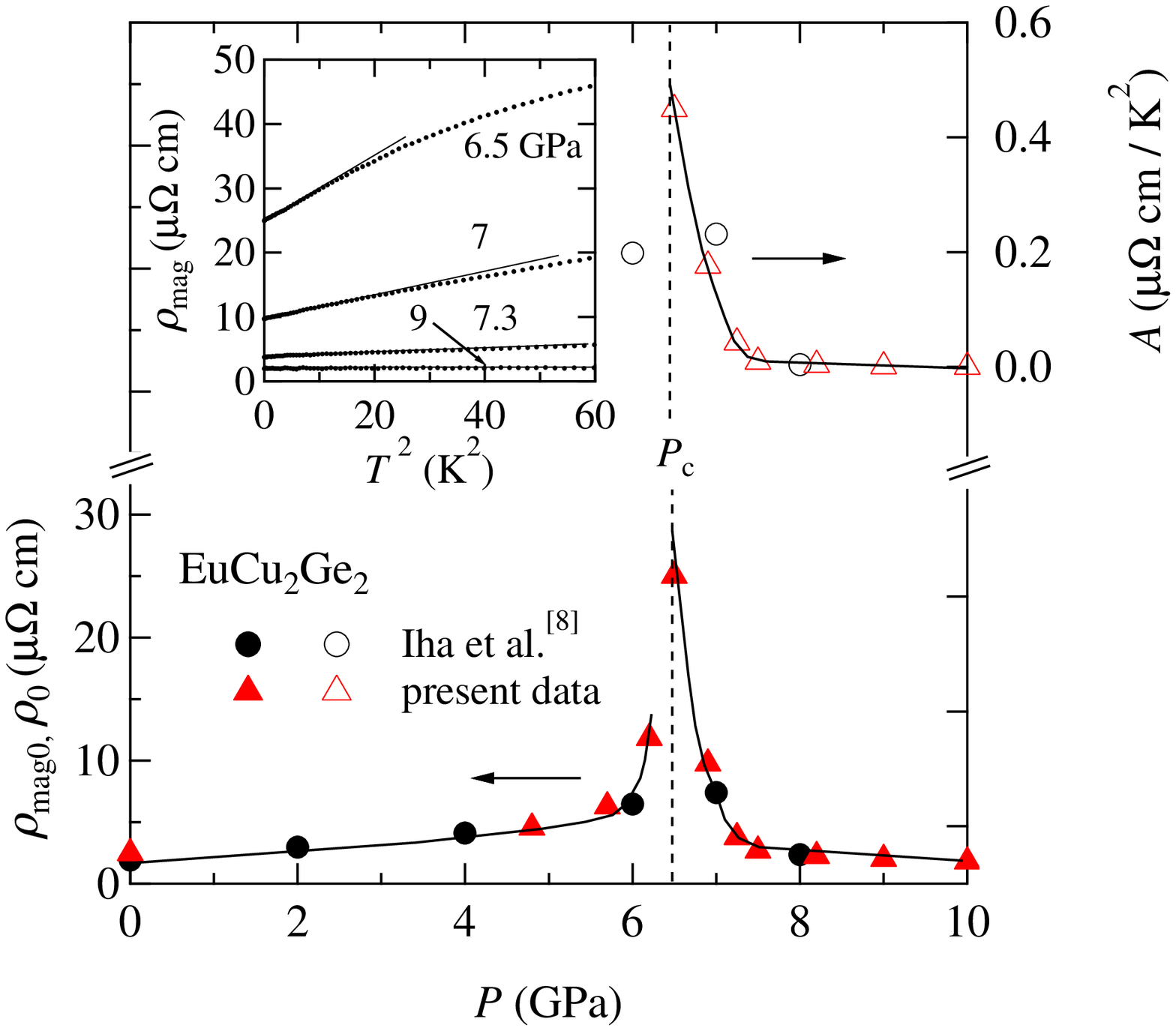}
\caption{\label{fig3} (color online) Pressure dependences of the coefficient of the $T^2$ dependence of the electrical resistivity $A$, residual resistivity $\rho_\mathrm{0}$, and magnetic residual resistivity $\rho_\mathrm{mag0}$. The solid lines serve as a visual guide. The circular and triangular symbols indicate our previous and present data, respectively. The open and closed symbols indicate the coefficient $A$ and $\rho_\mathrm{0}$ or $\rho$$_\mathrm{mag0}$, respectively. Note that the closed circles and triangles indicate residual resistivity $\rho_\mathrm{0}$ and $\rho_\mathrm{mag0}$, respectively. The inset shows the  $T^2$ dependence of $\rho$$_\mathrm{mag}$ at low temperatures.}
\end{center}
\end{figure}

From these data, we present the pressure $P$ $vs$ N$\mathrm{\acute{e}}$el temperature $T_\mathrm{N}$ phase diagram of EuCu$_2$Ge$_2$ in Fig. 2. The N$\mathrm{\acute{e}}$el temperature $T$$_\mathrm {N}$ becomes a maximum at $P = $ 6.2 GPa and drops to zero very sharply at 6.5 GPa. In the previous high-pressure study of Eu$_{2}$Ni$_{3}$Ge$_{5}$\cite{1}, the N$\mathrm{\acute{e}}$el temperature was found to increase to 7 GPa but to decrease continuously with further increasing pressure, suggesting that the valence changed continuously as a function of pressure. The  Kondo-like behavior of the electrical resistivity of EuCu$_2$Ge$_2$ at $T > $ $T_\mathrm {max}$ suggests the existence of a heavy-fermion state at low temperatures above $P$$_\mathrm {c}$. In the present study, on the other hand, the N$\mathrm{\acute{e}}$el temperature  became zero abruptly at around 6.5 GPa. This suggests that the valence of Eu increased continuously up to $P =\, $ 6.2 GPa, and then the quantum critical point (QCP) of the valence transition or sharp crossover of the valence arises abruptly at around $P=$ 6.5 GPa.

The theoretical understanding of Ce and Yb compounds has already been discussed in Refs. 16 and 17.  Indeed, this phenomenon of EuCu$_2$Ge$_2$ resembles a pressure effect on the prototypical heavy-fermion compound CeRhIn$_{5}$ in which the AF order disappears suddenly at $P =$ $P$$_\mathrm {c}$ $\simeq$ 2.4 GPa under a magnetic field of $H>$ 4 T,\cite{21, 22} which is associated with the Fermi surface change from a small to a large one as observed in a de Haas-van Alphen experiment\cite{23}. These behaviors have been understood theoretically in a unified fashion as a phenomenon that the sharp crossover or the criticality of the valence change cuts the AF order discontinuously in the case where the strength of the $c$-$f$ hybridization is relatively small\cite{20, 24}.

We consider here the Fermi liquid relation $\rho_\mathrm{mag} = \rho_\mathrm{mag0} + AT^{2}$ where the antiferromagnetic ordering disappeared. The magnetic resistivity $\rho_\mathrm{mag}$ is estimated by subtracting the electrical resistivity of YCu$_2$Si$_2$ (phonon part) \cite{18} from that of EuCu$_2$Ge$_2$ as $\rho_\mathrm{mag}$ = $\rho$(EuCu$_2$Ge$_2$) - $\rho$(YCu$_2$Si$_2)$ in order to remove the phonon contribution. $\rho_\mathrm{mag0}$ was determined by the magnetic residual resistivity at the lowest temperature of measurement, i.e., 10 mK. Figure 3 shows the pressure dependences of $\rho_\mathrm{mag0}$ and the coefficient $A$ assuming that the Fermi liquid relation of $\rho_\mathrm{mag} = \rho_\mathrm{mag0} + AT^{2}$ holds in the paramagnetic state. The Fermi liquid relation that persists up to about 1.8 K at 6.5 GPa spreads to a higher-temperature range as pressure is increased, as shown in the inset of Fig. 3. Correspondingly, $\rho_\mathrm{mag0}$ increases moderately as a function of pressure up to $P =\,$ 6 GPa. However, $\rho_\mathrm{mag0}$ increases steeply toward the sharp and huge peak at $P =$ $P$$_\mathrm {c}$ $\simeq$ 6.5 GPa, as shown in the lower part of Fig. 3. This supports the fact that this critical pressure $P$$_\mathrm {c}$ is related with the QCP of the valence transition as discussed in Ref. 22. The values of $A$ were obtained  assuming the Fermi liquid relation. The obtained $A$, which is proportional to the enhanced effective electron mass, decreases substantially above $P$$_\mathrm {c}$. The present $A$ value is 0.45 $\mu$$\Omega$ cm K$^{-2}$ at 6.5 GPa and corresponds to 510 mJ/mol$\cdot$K$^{2}$ if we assume the generalized Kadowaki-Woods relation as mentioned above. This  implies that the effective mass of quasiparticles is highly enhanced according to the generalized Kadowaki-Woods relation\cite{28, 29} or the temperature dependence of $\rho (T)$ changes from $T^{2}$ to $T^{n}$ with a smaller exponent of $n < 2$ due to the quantum criticality in one form or another.

\begin{figure}
\begin{center}
\includegraphics*[width=0.9\linewidth]{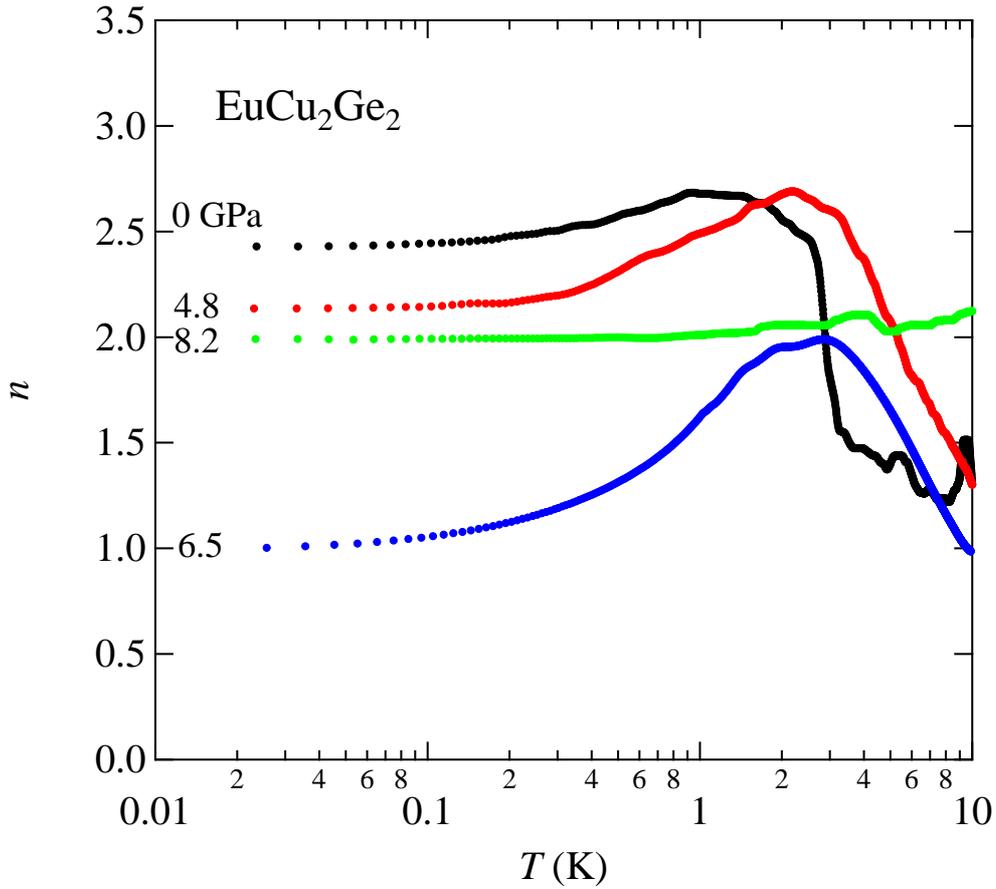}
\caption{\label{fig4} (Color online)  Power-law behavior for selected resistivity curves on a semilog scale.}
\end{center}
\end{figure}
Next, we focus on the pressure effect on the exponent $n$ of the power law dependence of the resistivity: $\rho_\mathrm{mag} = \rho_\mathrm{mag0} + BT^{n}$. The fit of the power law  $\rho_\mathrm{mag} = \rho_\mathrm{mag0} + BT^{n}$ to the resistivity data provides valuable information on the pressure dependences of the exponent $n$ of EuCu$_2$Ge$_2$. It is predicted that $n$ reaches 1.5 for the AF-QCP\cite{30}, while it reaches 1 for the QCP of the valence transition\cite{31}. In the present study, AF ordering abruptly disappears at 6.5 GPa, which is consistent with the change in the valence of Eu from nearly divalent toward the trivalent state. Indeed, there is evidence for the quantum valence fluctuations as discussed below.

Figure 4 shows the temperature dependence of the power-law behavior for selected resistivity curves, where the exponent $n$ is defined as
\begin{equation}
n \equiv \frac{\partial\mathrm{log}(\rho_\mathrm{mag} - \rho_\mathrm{mag0})}{\partial \mathrm{log}T}.
\end{equation}
In this analysis, $\rho_\mathrm{mag0}$ was used for the same value as shown in Fig. 3. Figure 5 shows the pressure dependence of exponent $n$. In the magnetic region (0 $<$ $P$ $<$ 6.2 GPa), $n$ is larger than the Fermi liquid value $n =$ 2, probably due to electron-magnon scattering. Approaching $P_\mathrm {c}$, the exponent $n$ decreases as a function of pressure and approaches $n =$ 1 slightly above 6.5 GPa, suggesting that $P_\mathrm {c}$ coincides with $P_\mathrm {v}$, the critical pressure for the quantum valence transition. Because $n$ is markedly changed, even with a slight change in pressure, the tuning of pressure around $P$$_\mathrm {c}$ is very difficult in experiments. Since the exponent $n$ appreciably deviates from 1.5 at $P = $ 6.5 GPa, the scenario based on critical AF fluctuations\cite{30} cannot  be applied to the present case but suggests that $P_\mathrm {c}$ $( = P_\mathrm {v}$) is located at slightly higher than $P = $ 6.5 GPa, which is consistent with the huge peak of $\rho_\mathrm{mag0}$ at $P = $ 6.5 GPa as discussed in Refs. 22 and 27.   Namely, the scenario based on critical valence fluctuations explains a gross feature of pressure effect on $n$ and $\rho_\mathrm{mag0}$ of EuCu$_{2}$Ge$_{2}$ observed near the critical pressure $P$$_\mathrm {c} \simeq 6.5$ GPa. Similar behavior was observed also in CeCu$_2$Ge$_2$ \cite{33}and CeRhIn$_{5}$\cite{21}. In the case of CeCu$_2$Ge$_2$\cite{33}, the exponent $n$ decreases rapidly with increasing pressure in the antiferromagnetic region and crosses almost exactly the value of $n$ = 1 at $P = $$P_\mathrm {c}$, and the $P$-interval where $n$ is smaller than two corresponds approximately to that of superconductivity. Also, in the case of CeRhIn$_{5}$, the exponent of $n$ approaches $n =$ 1 near $P =$ $P$$_\mathrm {c}$\cite{21, 22}. This is a signature of critical valence fluctuations. In the present pressure experiment, we did not observe superconductivity in EuCu$_2$Ge$_2$, unlike the cases of CeCu$_2$Ge$_2$ and CeCu$_2$Si$_2$ in which the superconducting transition temperature is enhanced at around $P$$_\mathrm {v}$\cite{32, 33}.

\begin{figure}
\begin{center}
\includegraphics*[width=0.8\linewidth]{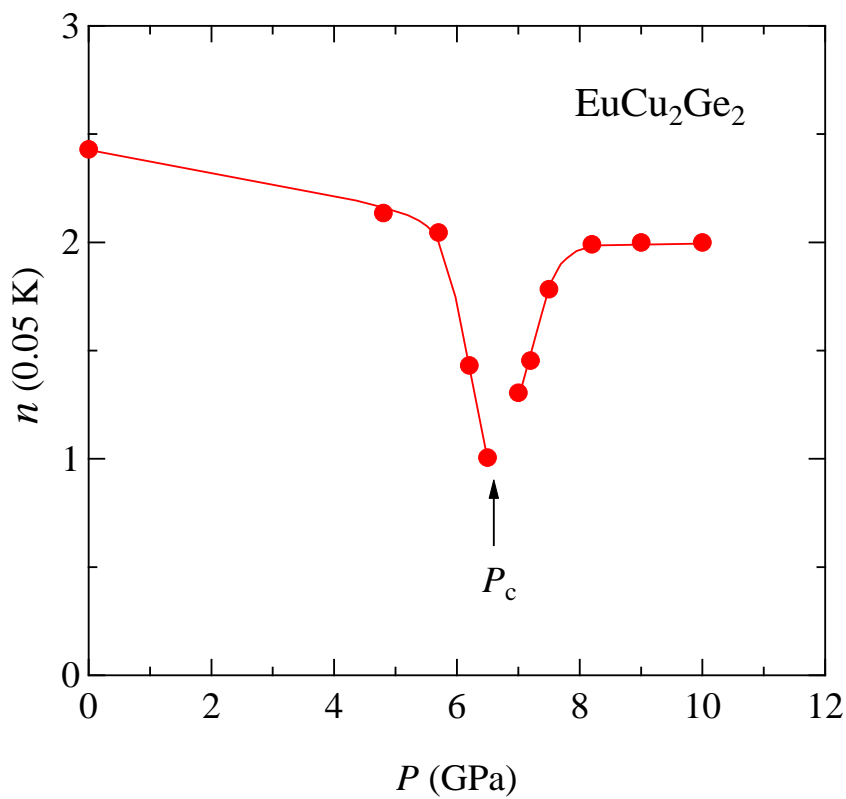}
\caption{\label{fig4} (Color online) Pressure dependence of exponent $n$ of the power law $\rho_\mathrm{mag} = \rho_\mathrm{mag0} + BT^{n}$ of the magnetic resistivity at 0.05 K for EuCu$_2$Ge$_2$.  The solid curves are guides to the eye.}
\end{center}
\end{figure}

In summary, we have studied the pressure effect on the electronic state of the AF compound EuCu$_2$Ge$_2$. The N$\mathrm{\acute{e}}$el temperature substantially increases from 15 K as a function of pressure and becomes a maximum of $T_\mathrm{N} =$ 27 K at 6.2 GPa. The resistivity at 6.5 GPa indicates no clear magnetic transition. In the present system, the valence of Eu changed continuously with increasing pressure up to 6.2 GPa, following the Doniach phase diagram. At $P$$_\mathrm {c}$ $\simeq$ 6.5 GPa, the N$\mathrm{\acute{e}}$el temperature abruptly dropped to zero, suggesting that a sharp or critical valence change occurs there from nearly a divalent state to one with a trivalent component of Eu$^{3+}$. The exponent $n$ obtained from the power law $\rho_\mathrm{mag} = \rho_\mathrm{mag0} + BT^{n}$ is very close to $n = $ 1 at $P = $ 6.5 GPa. These results indicate that  the case occurs where the pressure just above $P = 6.5$ GPa coincides with $P$$_\mathrm {v}$, corresponding to the quantum criticality of the valence transition.  This is the first observation of this phenomenon among the Eu compounds. The electronic specific heat coefficient, estimated from the generalized Kadowaki-Woods plot, was about 510 mJ/mol$\cdot$K$^{2}$, suggesting that the effective mass of the quasiparticles is highly enhanced around $P$$_\mathrm {c}$.\\

\section*{Acknowledgements}
We thank S. Nagasaki for helping with our high-pressure experiments. This work was partially supported by JSPS KAKENHI Grant Numbers JP19H00648, JP18H04329, JP17K05547, JP17K05555, and JP16K05453.

\end{document}